\documentstyle[preprint,aps]{revtex}
\begin{document}

\title{Spin-Charge Separation in Two Dimensions: A Numerical Study}
\author{Mihir Arjunwadkar, P. V. Panat and D. G. Kanhere \cite{email}}
\address{Department of Physics, University of Poona, Pune 411 007, India}
\date{}
\maketitle

\begin{abstract}
The question of spin-charge separation in two-dimensional lattices has
been addressed by numerical simulations of the motion of one hole in a
half-filled band. The calculations have been performed on finite clusters
with Hubbard and $t$-$J$ models. By comparing the time evolution of spin
and charge polarisation currents in one and two dimensions, evidence
in favor of spin-charge separation in two dimensions is presented.
In contrast with this, spin-charge separation is absent in a
highly doped, metallic, system.
\end{abstract}

\pacs{PACS number(s): 71.27.+a,74.20.Mn}

The crucial issue in understanding the physics of strongly-correlated
systems is the nature of the ground state and low-lying excitation
the normal state of doped high-$T_c$ materials. In particular,
intense attention has been focused on the question of spin-charge
separation and non-Fermi (or Luttinger) liquid behaviour of Hubbard
and related models in 2D. It has been forcefully argued by Anderson
\cite{pwa} that the breakdown of Fermi liquid behaviour and the
phenomenon of spin-charge separation, well established in 1D, is
carried over to 2D systems and weakly coupled chains and planes as
well. These arguments are substantiated by physically motivated
scattering theory and anomalous fermionic backflow \cite{baskaran}
calculations. The interlayer tunneling mechanism proposed by Wheatley,
Hsu, and Anderson \cite{wha1,wha2} crucially depends on the existence of
spin-charge decoupled layers coupled via a weak interlayer hopping term.
However, in spite of extensive investigations, both analytical and numerical,
so far there is no clear, convincing signature of non-Fermi liquid behaviour
in 2D.

The situation in 1D is clearer due to the exact solutions of Hubbard
\cite{shiba} and Tomonaga-Luttinger models and other calculations.
\cite{1D} The physics in 1D, in the absence of a magnetic field, is
characterized by three parameters, which are $u_\rho$ (charge velocity),
$u_\sigma$ (spin velocity), and $K_\rho$ (coefficient that determines the
long range decay of correlation functions). $u_\rho$ and $u_\sigma$,
in the large-$U$ limit, are, respectively, given by
$2t\sin \pi n$ and $\frac{2\pi t^2}{U} (1 - \frac{\sin 2\pi n}{2\pi n})$
where $t,U$ are the Hubbard model parameters and $n$ is the particle
density ($n \leq 1$). This physics is described in terms of spinons
(excitations carrying spin-$1/2$ but no charge) and holons (excitations
carrying unit charge but no spin), and implies that the two kinds of
excitations have altogether different dynamics. This is called
spin-charge separation.

Thus in the case of 1D, the non-Fermi liquid behaviour typically
manifests itself as (i) the power-law behaviour of correlation functions;
in particular, of the momentum distribution around ${\bf k}_F$, and
(ii) spin-charge separation. Away from 1D, the only exact results
available are due to Fabrizio and Parola \cite{parola} on coupled chains
with a modified Tomonaga-Luttinger model, in which case spin-charge
separation is shown to exist. On the other hand, calculations which
{\em do not} observe spin-charge separation in coupled chains, are
quite numerous. \cite{schulz} Numerical attempts, in 2D, as regards
the power-law behaviour of the momentum distribution are inconclusive.
\cite{nK} In this work, we prefer to focus our attention on the
possibility of observing spin-charge separation by simulating the motion
of one hole in a half-filled band. Indeed, we do demonstrate spin-charge
separation in 1D, and by comparing the behaviour seen in 2D with that
in 1D, present a favourable evidence for such a separation in 2D.

Now we describe the simulation procedure. We introduce a Gaussian
hole, at time $t=0$, into the ground state $|G\rangle$ of a cluster,
Hubbard or $t$-$J$, at half-filling, obtained by exact
diagonalization. \cite{wha2,exact}
The resulting state can be written as
\begin{equation}
|\psi (0)\rangle = \sum_{i} e^{i {\bf K}_0 . ({\bf r}_i - {\bf R}_0) -
             \beta |{\bf r}_i - {\bf R}_0|^2} c_{i\sigma} |G\rangle.
\end{equation}
The charge distribution at $t=0$ is centered around ${\bf K}_0$ in
momentum space with spread $\sim \sqrt{\beta}$ and ${\bf R}_0$
in real space with spread $\sim \frac{1}{\sqrt{\beta}}$. This state
is then subjected to (second-order) time evolution
\begin{equation}
|\psi(t+\Delta t)\rangle = (1 - i\Delta tH - {1 \over 2}(\Delta tH)^2)
                           |\psi(t)\rangle,
\end{equation}
during which various quantities pertaining to charge and spin are
computed. It is well known that the finite time step (and the approximate
form for the time evolution operator) tends to make the evolution
nonconservative. This error can only be reduced by using a small enough
time step and by including higher-order terms in the time evolution
operator, which is the reason for using the second-order form rather than
the (numerically less costly) first-order one. Further, the time step is
chosen to ensure that the energy expectation value in the state
$|\psi(t)\rangle$
always remains within a few percents of its initial value. This procedure
is indeed the same as that of Jagla {\it{et al.}}, \cite{jagla} except that
we are using the second-order approximation to the time evolution
operator. They demonstrated that spin-charge separation can readily be
observed on small 1D clusters by numerical means. They failed to see it
in 2D because of very high doping. This point will be discussed later.
The ``visual'' results of their paper in 1D have been confirmed by us.

In order to find an appropriate quantity that would reflect spin-charge
separation, let us recall that spin-charge separation results from the
dynamical independence of spin and charge degrees of freedom, which
should be reflected in the dynamical behaviour of charge ($+$) and spin
($-$) densities:
\begin{equation}
\rho_\pm (i,t) =
\frac {\langle\psi(t)|(n_{i\uparrow} \pm n_{i\downarrow})|\psi(t)\rangle -
B_\pm}
      {\sum_{i} (\langle\psi(0)|(n_{i\uparrow} \pm
n_{i\downarrow})|\psi(0)\rangle
       - B_\pm)}.
\end{equation}
Here $B_+$ is the average background charge (number of electrons per site,
$=1$ for the half-filled ground state) and $B_-$ is the average background
spin ($=0$ for the half-filled ground state belonging to the $S_z=0$
subspace). These are thus the densities associated with the doped particle,
and are normalized to unity. In the non-initeracting ($U=0$) case,
we expect $\rho_+ (i,t)$ to vary in time identically as $\rho_-(i,t)$,
whereas in the spin-charge decoupled case, these two quantities should
show a non-trivially different behaviour in time. It is more convenient
to define site-independent aggregate quantities from these, which are
polarisations \cite{mahan}
\begin{equation}
{\bf P}_\pm (t) = \sum_i {\bf r}_i \rho_\pm(i,t).
\end{equation}
Qualitatively, the polarisations reflect the center-of-mass movement of
the charge and spin peaks. Since they are origin-dependent, we prefer to
look at their rates of change, ${d {\bf P}_\pm (t) \over d t}$, which are
the average classical currents set up in the system due to the
time evolution of the hole packet, and are a measure of charge and spin
group velocities. \cite{note1}

We have carried out extensive simulations on the following 1D and 2D
clusters: (i) six-site Hubbard ring, (ii) ($4+4$) site coupled Hubbard
chains, (iii) $4 \times 4$ $t$-$J$ plane, with one hole doped in the
half-filled ground state, and (iv) $4 \times 4$ Hubbard plane with one
{\em extra electron} in the two-electron ($S_z=0$) ground state.
Periodic boundary conditions have been used in all the cases. Note that
the $t$-$J$ model has been used only in the case where it is impossible
to work with the Hubbard model due to the large-basis problem. This, we
think, is acceptable, since we believe that the essential physics of the
large-$U$ Hubbard model is contained in the $t-J$ model. Simulations are
done by varying the width of the Gaussian, $\beta$, and for all
${\bf K}_0$ appropriate for a given cluster, as well as the model
parameters $U$ or $J$.

We first present the results on the 1D six-site ring, for which
spin-charge separation is known to exist. These will bring out the
characteristic behaviour of currents ${d {\bf P}_\pm (t) \over d t}$ in the
spin-charge separated case as against the noninteracting, undecoupled
case. For all these runs, we choose $\beta=0.1$ and
${\bf K}_o={-2\pi \over 3}$. All energies and times are measured with
respect to the Hubbard parameter $t=1$. Figure 1 shows the currents as a
function of time (up to 2000 steps, $\Delta t = 0.01$) for values of
$U=0$ [Fig. 1(a)], $U=1$ [Fig. 1(b)] and $U=20$ [Fig. 1(c)].
As expected, for the noninteracting
case [Fig. 1(a)], the two curves overlap for all times, and show a
periodic behaviour characteristic of the free motion of a hole packet on
a periodic lattice. This also means that $\rho_\pm (i,t)$, at each site
$i$, oscillates with the same frequency. As $U$ increases [Fig. 1(b)],
the charge and spin currents start ``separating,'' indicating the response
of the background, although the periodicity of the $U=0$ case is still
evident, including the locations of the peaks. However, the behaviour of
${d {\bf P}_+ \over d t}$ and ${d {\bf P}_- \over d t}$ are dramatically
different from each other for large $U$, beyond $U \sim 5$, indicating
the decoupling of spin and charge dynamics [Fig. 1(c)]. The contrast
between Fig. 1(b) and 1(c) can be related to the ``stiffening'' of the
antiferromagnetic background with increasing $U$. In order to understand
the highly oscillatory behaviour of the currents, let us note that by
creating a hole at $t=0$ in the background (i.e., half-filled
ground state), we have created a state that is {\em not} an eigenstate
of the system, and can be written as a linear combination of ground and
excited states. In 1D, these excited states can always be described
\cite{1D} as spinon and holon excitations, having different dynamics,
and we believe that the observed behaviour of the currents in Fig. 1(c)
is a direct consequence of this.

Now we present the results for the 2D system ($4 \times 4$ $t$-$J$
cluster). For this case we choose
${\bf K}_0=({-\pi \over 2},{-\pi \over 2})$, $\beta=1$,
and $J=0.1$. The initial Gaussian hole packet has been placed symmetrically
with respect to the entire lattice, because of which the $y$ component
of the currents varies identically as the $x$ component, for this
${\bf K}_0$. \cite{note2} We thus display, in Fig. 2, only the
$x$ components of spin and charge currents. Clearly, the two currents
indicate different dynamics, in that, their magnitudes as well as
directions (and phases) are different from each other almost all the
time. This feature is qualitatively similar to Fig. 1(c) for the 1D
case, $U=20$. We interpret this as a signature of spin-charge separation
in 2D. Qualitatively similar behaviour is seen for $J$ as large as $0.8$,
which is not surprising, since the $t$-$J$ model has strong correlations
built-in because of the elimination of double occupancy, throughout the
parameter range.

It is to be noted that the above mentioned behaviour is observed for
{\em one} hole doped into the half-filled ground state of the $4 \times 4$
cluster. In order to show that this behaviour, indicative of spin-charge
separation, {\em does not} persist for high doping--the metallic case,
we examine the time evolution of one {\em extra electron} in the
two-electron ($S_z=0$) ground state of the $4 \times 4$ Hubbard
cluster as a function of $U$. We have ${\beta=1.0}$,
${\bf K}_0=({-\pi \over 2},0)$ and the gaussian packet is again placed
symmetrically with respect to the lattice. There is no current in the
$y$ direction because of this choice of ${\bf K}_0$ and ${\bf R}_0$
\cite{note2}. Fig. 3 (a) and (b) depict the $x$ currents for $U=0.1$ and
$U=10$ respectively. Indeed, it is observed that the familiar periodic
behaviour (with more or less overlapping curves), indicative of a
weakly interacting, spin-charge {\em undecoupled} system, is observed
even for $U$ as high as $10$. This is precisely the reason why Jagla
{\it et al.} \cite{jagla} did not observe spin-charge separation in
their study in 2D.

In conclusion, we have studied numerically the time evolution of one hole
in a half-filled band, and by comparing the behaviour of spin and charge
currents in 2D with that in 1D, we have presented evidence in favour
of spin-charge separation in 2D lattices. We have also demonstrated that
this phenomenon is observed only in the low-doped systems (one hole in
the present case) and is absent in the high-doping limit (low electron
densities). Finally, we note that our calculations on $4+4$ coupled chains
(not presented here) also display a similar behaviour which is indicative
of spin-charge separation in coupled systems.

\acknowledgements

It is a pleasure to acknowledge G. Baskaran, MatScience (Madras)
for a number of illuminating discussions, and A. D. Gangal for many
useful suggestions. One of us (MA) would like to acknowledge, in
addition, V. N. Muthukumar and Rahul Basu (MatScience) for extensive
discussions and encouragement, as well as the Council for Scientific
and Industrial Research (CSIR), New Delhi for financial support.
Partial financial assistance was provided by the Department
of Science and Technology (DST) under Project No. SP/S2/M-47/89.

\newpage

\narrowtext

{\bf Figure Captions}\\[1cm]

FIG. 1  Currents ${d {\bf P}_\pm (t) \over d t}$ as functions of
time for one hole in the half-filled ground state of the six-site
Hubbard ring for different values of $U$:  (a) $U=0$, (b) $U=1$, (c) $U=20$;
solid curve, charge current; dotted curve, spin current. \\[0.5cm]

FIG. 2  $x$ components of currents ${d {\bf P}_\pm (t) \over d t}$
as functions of time for one hole in the half-filled ground state of the
($4 \times 4$)-site $t$-$J$ cluster;
solid curve, charge current; dotted curve, spin current. \\[0.5cm]

FIG. 3  $x$ components of currents ${d {\bf P}_\pm (t) \over d t}$
as functions of time for one extra electron in the two-electron ground
state of the ($4 \times 4$)-site Hubbard cluster for different values of $U$:
(a) $U=0.1$, (b) $U=10$;
solid curve, charge current; dotted curve, spin current.

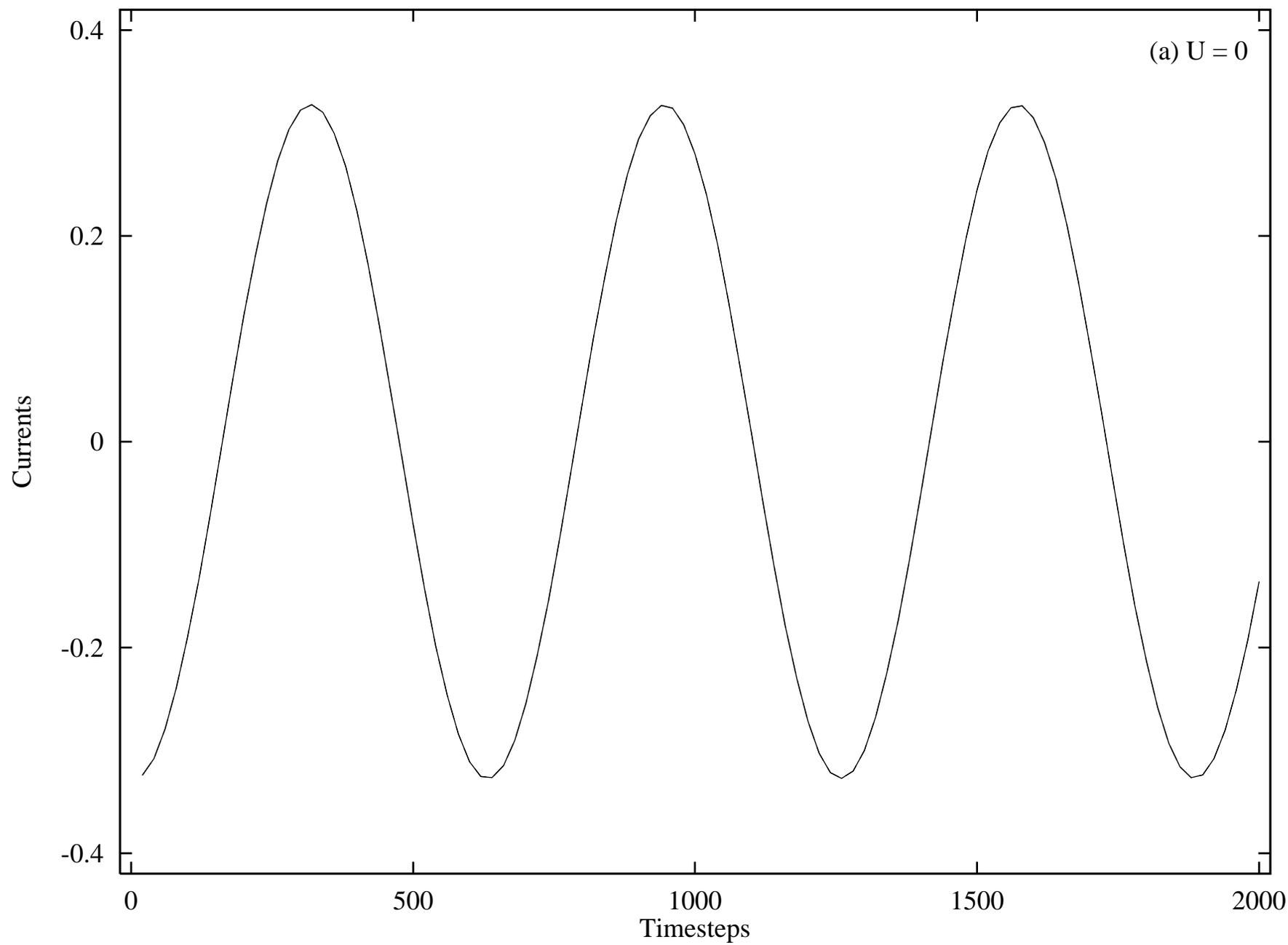
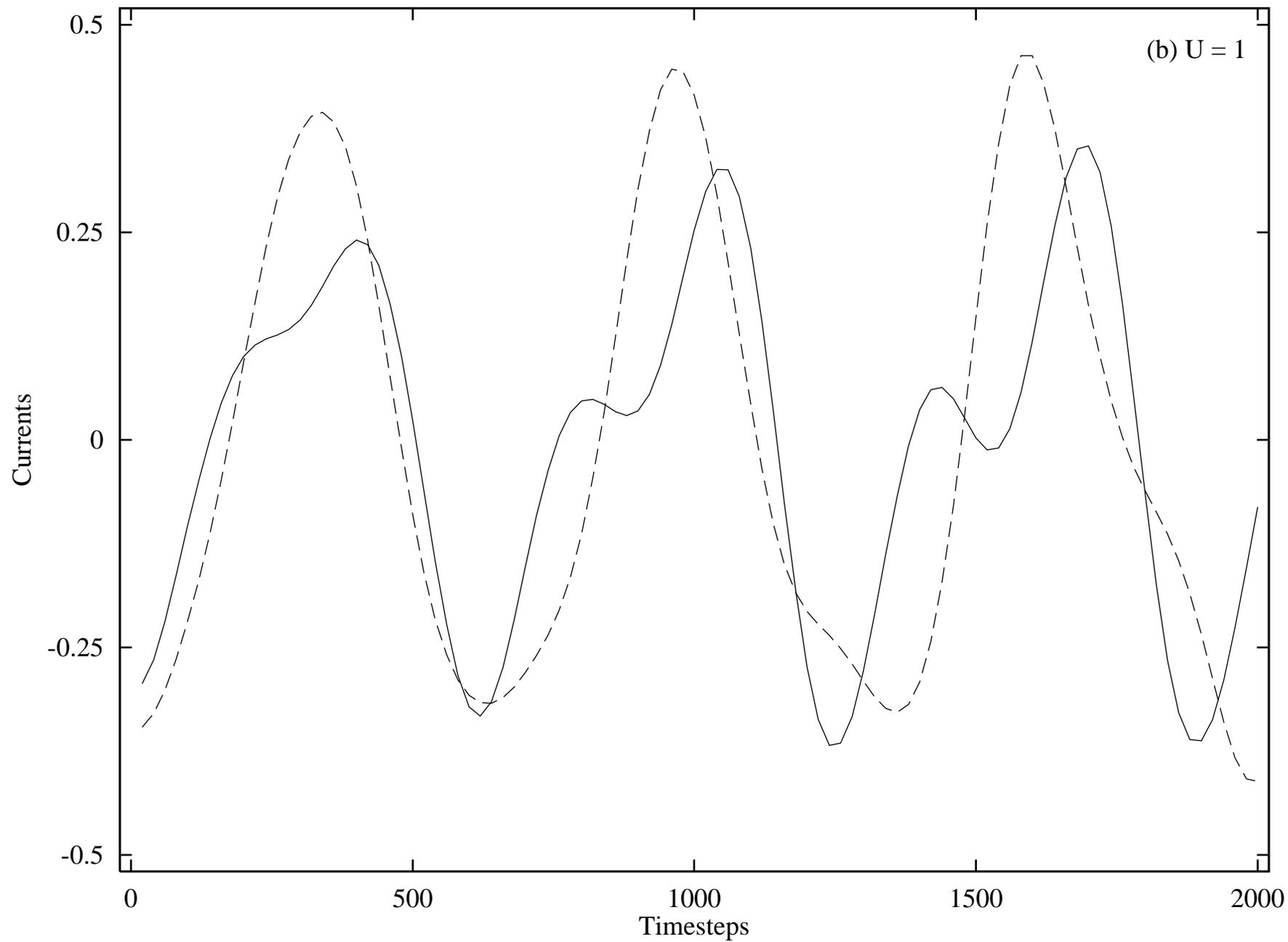
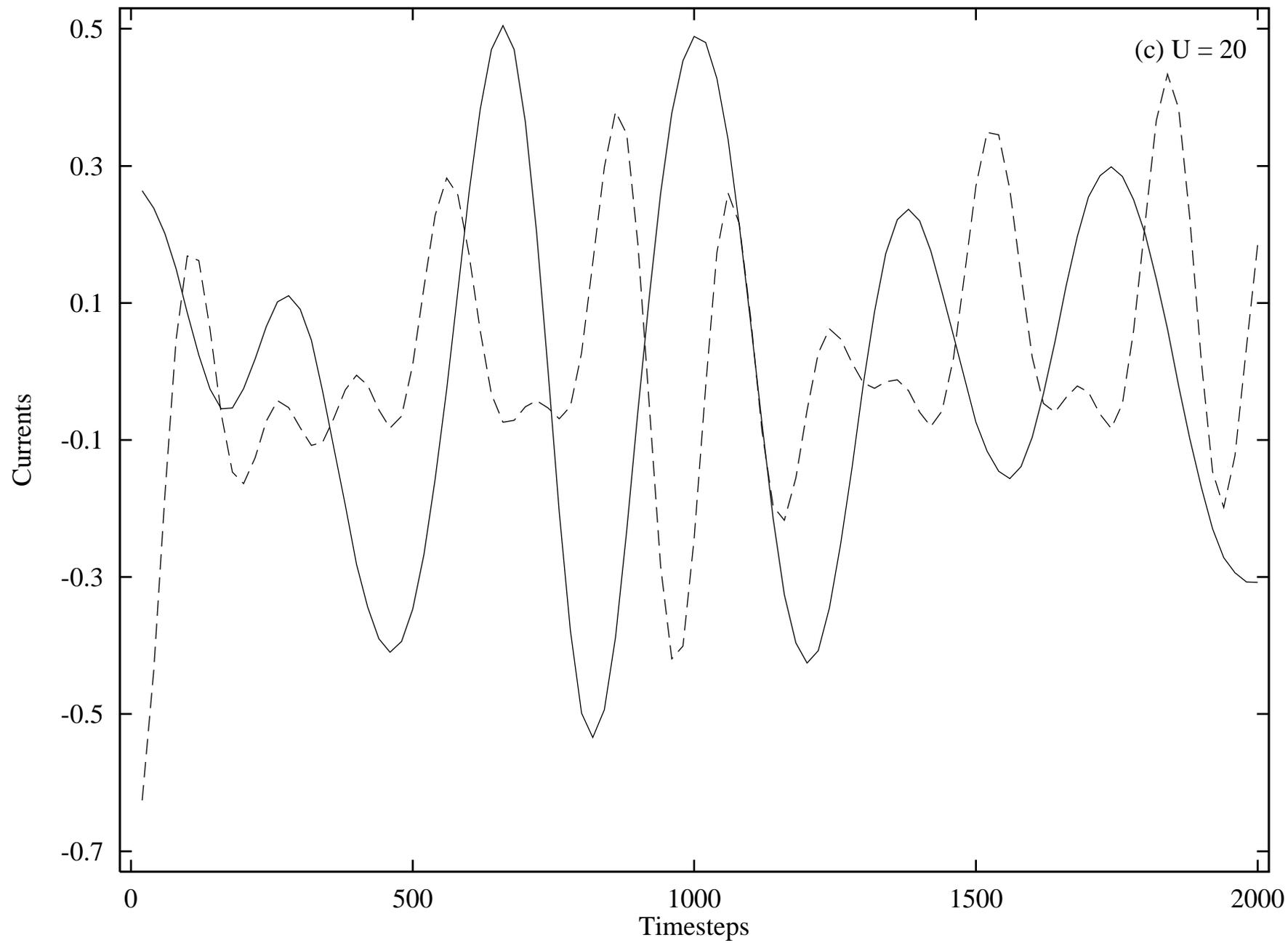
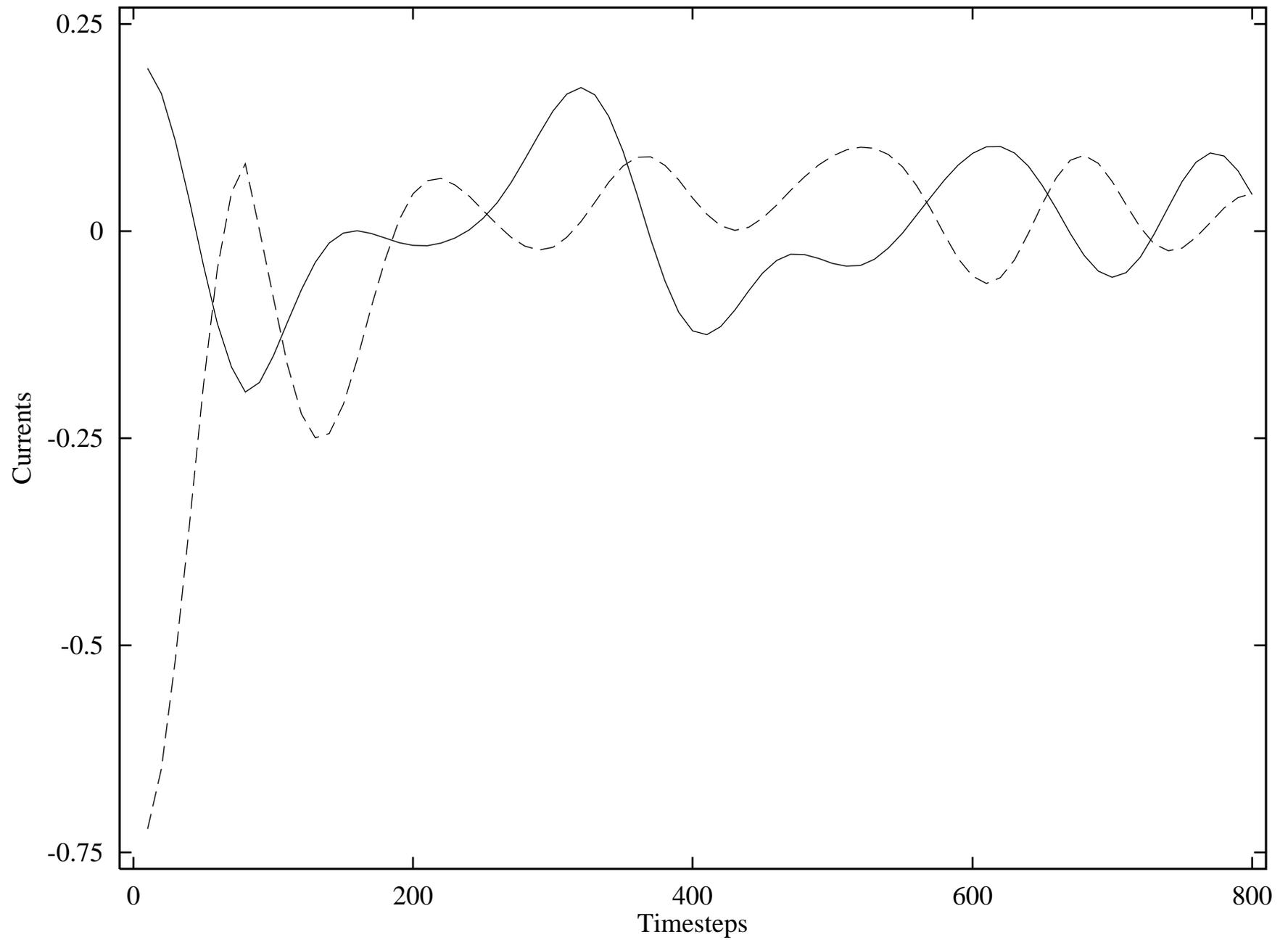
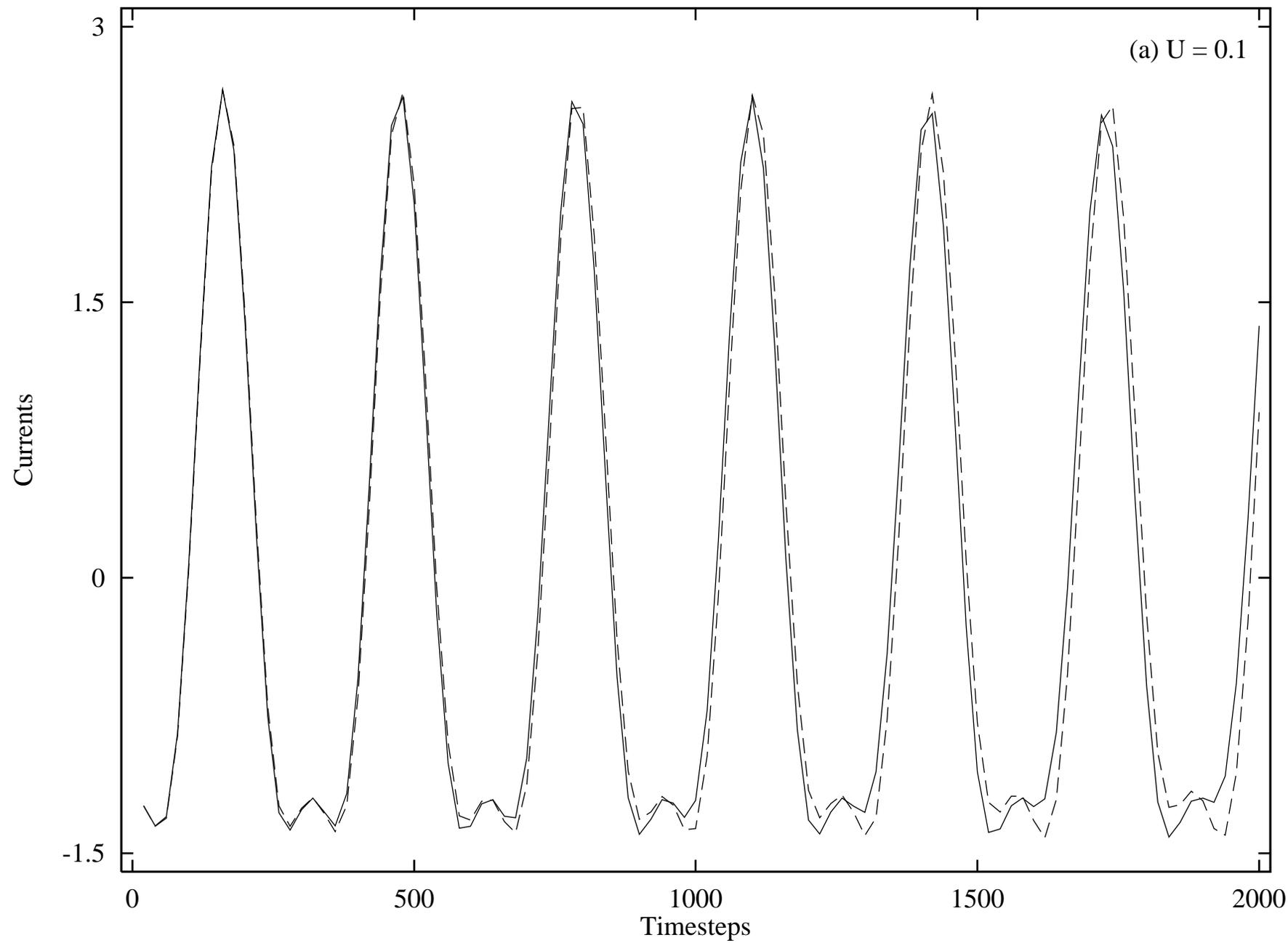
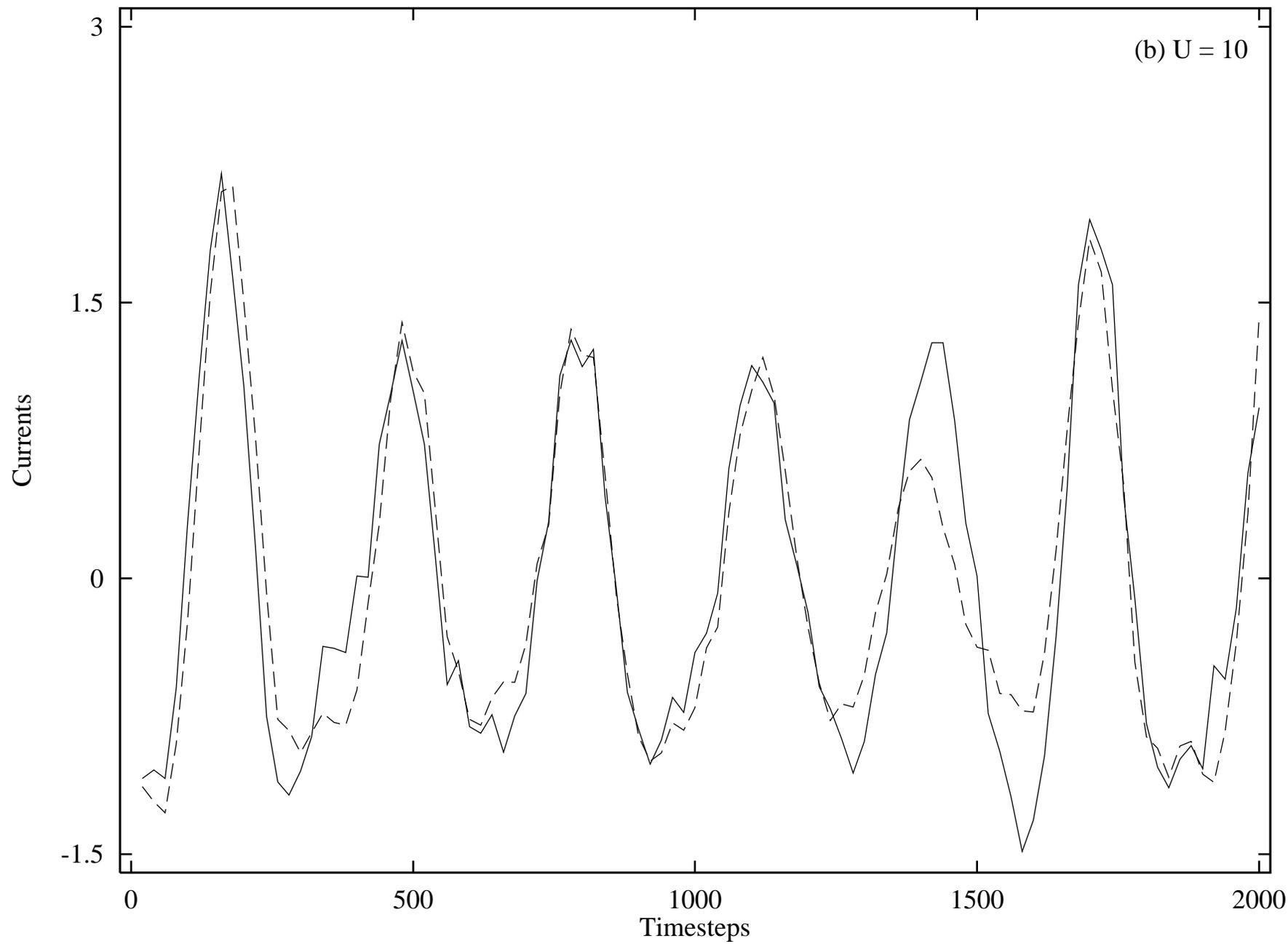
\end{document}